\documentclass[12pt]{article}
\usepackage{amsfonts}
\usepackage{amssymb}
\usepackage{latexsym}
\usepackage{graphicx}
\usepackage[english]{babel}
\begin{document}
\input epsf

\def\p{\partial}
\def\h{{1\over 2}}
\def\be{\begin{equation}}
\def\bea{\begin{eqnarray}}
\def\ee{\end{equation}}
\def\eea{\end{eqnarray}}
\def\d{\partial}
\def\la{\lambda}
\def\eps{\epsilon}
\def\bb{\bigskip}
\def\mm{\medskip}
\newcommand{\dm}{\begin{displaymath}}
\newcommand{\edm}{\end{displaymath}}
\renewcommand{\b}{\tilde{B}}
\newcommand{\gm}{\Gamma}
\newcommand{\ac}[2]{\ensuremath{\{ #1, #2 \}}}
\renewcommand{\ell}{l}
\newcommand{\z}{\ell}
\newcommand{\newsection}[1]{\section{#1} \setcounter{equation}{0}}
\def\bb{$\bullet$}
\def\Qbar{{\bar Q}_1}
\def\QPbar{{\bar Q}_p}

\def\q{\quad}

\def\bn{B_\circ}

\let\a=\alpha \let\b=\beta \let\g=\gamma \let\d=\delta \let\e=\epsilon
\let\c=\chi \let\th=\theta  \let\k=\kappa
\let\l=\lambda \let\m=\mu \let\n=\nu \let\x=\xi \let\r=\rho
\let\s=\sigma \let\t=\tau
\let\vp=\varphi \let\vep=\varepsilon
\let\w=\omega      \let\G=\Gamma \let\D=\Delta \let\Th=\Theta
                     \let\P=\Pi \let\S=\Sigma

\def\h{{1\over 2}}
\def\t{\tilde}
\def\r{\rightarrow}
\def\nn{\nonumber\\}
\let\bm=\bibitem
\def\Kt{{\tilde K}}
\def\b{\bigskip}

\let\p=\partial

\begin{flushright}
\end{flushright}
\vspace{20mm}
\begin{center}
{\LARGE  How fast can a black hole release its information?}
\\
\vspace{18mm}
{\bf  Samir D. Mathur\footnote{mathur@mps.ohio-state.edu} }\\

\vspace{8mm}
Department of Physics,\\ The Ohio State University,\\ Columbus,
OH 43210, USA\\
\vspace{4mm}
\end{center}
\vspace{10mm}
\thispagestyle{empty}
\begin{abstract}

\b

When a shell collapses through its horizon,  semiclassical physics suggests that information cannot escape from this horizon. One might hope that nonperturbative quantum gravity effects will change this situation and avoid the `information paradox'. We note that string theory has provided a set of states over which the wavefunction of the shell can spread, and that the number of these states is large enough that such a spreading would significantly modify the classically expected evolution. In this article we perform a simple estimate of the spreading time, showing that it is much shorter than the Hawking evaporation time for the hole. Thus information can emerge from the hole through the relaxation of the shell state into a linear combination of fuzzballs.

\end{abstract}
\vskip 1.0 true in

\newpage
\setcounter{page}{1}

\section{The information paradox}

Consider a shell of matter that collapses to form a black hole. As the shell passes through its horizon, the light cones `tip over' so that any particle inside the horizon is forced to move towards the center of the hole, ending its trajectory at the singularity.
As a consequence the region near the horizon becomes the `vacuum' in this classical picture; any matter near this horizon either flows off to infinity or gets sucked inside the hole.

Quantum effects cause the black hole to slowly leak away energy by the creation of particle-antiparticle pairs at the horizon. But since these pairs are created out the vacuum, the emerging quanta carry no information about the the matter which made the hole. Thus in this semiclassical picture we get a loss of unitarity -- the well known black hole information paradox \cite{hawking}.

One might hope that nonperturbative quantum gravity effects like `tunneling' will resolve this paradox. But for this to happen, we needs three things

\b

(a) A set of states to tunnel {\it to}.

\b

(b) The probability of tunneling to these states should be high enough; since the black hole looks like a macroscopic, smooth classical object
it may seem that the probability to tunnel to something else will be small, and classical physics should be accurate.

\b

(c) Even if we have (a), (b), we still need to know that this tunneling will happen in a time shorter than the Hawking evaporation time of the hole
\be
t_{tunnel}\ll t_{evap}
\label{basic}
\ee
Otherwise the nonperturbative process will not help solve the information question -- the state of the created pairs will be essentially the same as the one in Hawking's semiclassical computation.

\bigskip

In this article we will recall how string theory has provided us with (a) and (b), and we will then perform a simple estimate and observe that we naturally arrive at (c).

\section{Black hole microstates}

People have long sought to construct `hair' on black holes to account for the large entropy \cite{bek}
\be
S_{bek}={A\over 4G}
\ee
But examining perturbations to the hole geometry gave no such hair. It turns out that the hair are {\it nonperturbative} deformations of the geometry. String theory leads to the realization that 6 additional compact dimensions beyond the visible 3+1 are needed to get a consistent theory of gravity. These additional directions give gauge fields by Kaluza-Klein (KK) reduction, and it is known that no perturbative hair can be found for gauge fields either. But now consider the following {\it nonperurbative} deformation. Take one compact circle $S^1$, and use it to make a KK-monopole at some location $\vec r_1$ in the non-compact space. This means that the $S^1$ fibers nontrivially over the $\vec r$ space, so  the construction uses the noncompact directions in an  essentially nonperturbative manner. At some other location $\vec r_2$ there will be an anti-KK-monopole, so that we generate no net KK charge. 

This geometry will have no horizon or singularity and might look nothing like the black hole that we wanted to consider. But it has the same mass and charges as the hole, and in string theory one can show there will be one subfamily of black hole microstates that will have this form. More complicated states can be made using more KK pairs, till we reach situations where the monopoles are so close together than quantum effects must be considered, and quantum  stringy fuzz fills up an entire ball  shaped region. The fuzzball conjecture says that different states of this kind account for all the states of the black hole; none of these states have a traditional horizon which would have the vacuum in its vicinity -- the black hole microstates are horizon size `fuzzballs' that can radiate like any other object that emits (information carrying) radiation from its surface \cite{fuzzballs}.

Now suppose we have a shell that is collapsing through its horizon. The fuzzball states form a complete set of energy eigenstates for the system, so this collapsing shell state can be written as a superposition of fuzzball states (which we label by their energies $E_k$) 
\be
|\psi\rangle=\sum_k c_k |E_k\rangle
\label{sum}
\ee
Can the wavefunction of the shell spread over this entire space of states? Let us first estimate a `tunneling amplitude' ${\cal A}\sim e^{S_{tunnel}}$ between the shell state and any one of the fuzzball states. In \cite{essay} it was argued that
\be
S_{tunnel}\sim {1\over G}\int \sqrt{-g} R =\alpha GM^2, ~~~\alpha=O(1)
\label{six}
\ee
where we have used the black hole length scale $GM$ to estimate the curvature scale in $S_{tunnel}$. 
This gives
\be
{\cal A}\sim e^{-S_{tunnel}}\sim e^{-\alpha GM^2}, ~~~\alpha=O(1)
\label{four}
\ee
This is, as expected, a very {\it small} number, but on the other hand the {\it number} of states that we can tunnel to is
\be
{\cal N}\sim e^{S_{bek}}\sim e^{GM^2}, 
\label{three}
\ee
a  very {\it large} number. We now see that (\ref{three}) can offset (\ref{four}), making a spreading over all fuzzball states possible. Thus while our first impression suggests that a black hole is a classical system, this is in fact not true; the enormous entropy of the hole gives a very large phase space of states, and even though there is a very small probability to tunnel to {\it one} of these states, the probability can be order unity to tunnel to {\it some} linear combination of the fuzzball states.

\section{The time for tunneling}

Our question now is: can we argue that this spread will happen over a time less than the Hawking evaporation time? Recall how we compute the tunneling time in a double-well potential. The wavefunction of a quantum in the left well can be written as a superposition of symmetric and antisymmetric wavefunctions
\be
|\psi\rangle_L={1\over \sqrt{2}}|\psi\rangle_S+{1\over \sqrt{2}}|\psi\rangle_A
\ee
These eigenfunctions $ |\psi\rangle_S, |\psi\rangle_A$ have slightly different energies, so that the total state evolves as
\be
|\psi\rangle_L={1\over \sqrt{2}}e^{-iE_S t}|\psi\rangle_S+{1\over \sqrt{2}}e^{-iE_A t}|\psi\rangle_A
\ee
After a time
\be
t_{dephase}=t_{tunneling}= {\pi\over E_A-E_S}\equiv {\pi\over \Delta E}
\ee
the  $S,A$ parts of the wavefunction are out of phase, and the wavefunction has tunneled to the right well state
\be
|\psi\rangle_R={1\over \sqrt{2}}|\psi\rangle_S-{1\over \sqrt{2}}|\psi\rangle_A
\ee

Now let us estimate  $t_{dephase}$ for our black hole problem. First,  note that  to make the shell collapse to a black hole we must localize the matter in the shell so that it fits in a radius $\ll R$, where $R$ is the Schwarzschild radius for the shell.  This needs a spread in radial momentum 
\be
\Delta P\gg  {1\over R}
\ee
 The energy of the shell is $E\sim {P^2\over 2M}$, and so the uncertainty in $E$ will be
\be
\Delta E\sim {P\Delta P\over M}\gg {(\Delta P)^2\over M}\gg {1\over MR^2}
\ee
Thus
  \be
t_{dephase}\sim {1\over \Delta E}\ll MR^2
\ee   
But the Hawking evaporation time for a Schwarzschild hole (in all dimensions) is
\be
t_{evap}\sim { MR^2}
\ee
Thus we find that the time over which the the wavefunction `dephases to fuzzballs' is shorter than the Hawking evaporation time
\be
t_{dephase} \ll t_{evap}
\ee

 \section{Summary}
 
 We thus have a complete path to understanding the resolution of the black hole information paradox. The two main peculiarities of black holes are (a) the very large entropy, which is large because of a power of ${1\over \hbar}$ and (b) the very long time of Hawking evaporation, which again has a ${1\over \hbar}$ when measures in units of the black hole radius. We now see that even though the black hole looks like a classical object, its behavior need {\it not} be classical at all. In the path integral formulation of quantum mechanics we have a classical action, and a phase space measure. For macroscopic objects, the measure factor is typically ignorable for the evolution. But since the black hole has such a large entropy, it has a very large phase space. The spread of the wavefunction over this phase space competes with the contribution of the classical action, and makes the physics essentially {\it quantum}.  Of course given enough time the wavefunction for {\it any} system spreads; thus we need to check that the timescale for spreading in our case is shorter than the physical timescale of interest -- the Hawking evaporation time. This is what we indeed found in our simple estimate.
 
 Thus the collapsing shell changes to a linear combination of fuzzball states in a time shorter than the Hawking evaporation time, and since these fuzzball states do not have `horizons', information can emerge just as from a piece of burning coal. It would be interesting to compare the estimates here with general ideas about information release times discussed in \cite{susskind}.

\section*{Acknowledgements}

This work was supported in part by DOE grant DE-FG02-91ER-40690.

\end{document}